\documentclass[twocolumn]{aastex631} 
\usepackage[version=3]{mhchem}
\usepackage{comment}

\newcommand{\mum}{$\rm \mu \mathrm{m}~$}

\newcommand{\OIII}{[\textsc{Oiii}]\,}
\newcommand{\OII}{[\textsc{Oii}]\,}
\newcommand{\HII}{H{\sc ii}~}

\usepackage{color}

\newcommand{\notice}[1]{\textcolor{black}{#1}}

\shorttitle{Clumpy galaxies in zoom-in simulations}
\shortauthors{Nakazato et al.}

\graphicspath{{./}{figures/}}
\newcommand{\figdir}{./}

\begin{document}

\title{A merger-driven scenario for clumpy galaxy formation in the epoch of reionization: Physical properties of clumps in the FirstLight simulation}

\correspondingauthor{Yurina Nakazato}
\email{yurina.nakazato@phys.s.u-tokyo.ac.jp}
\author[0000-0002-0984-7713]{Yurina Nakazato}
\affiliation{Department of Physics, The University of Tokyo, 7-3-1 Hongo, Bunkyo, Tokyo 113-0033, Japan}

\author[0000-0002-8680-248X]{Daniel Ceverino}
\affiliation{Universidad Autonoma de Madrid, Ciudad Universitaria de Cantoblanco, E-28049 Madrid, Spain}
\affiliation{CIAFF, Facultad de Ciencias, Universidad Autonoma de Madrid, E-28049 Madrid, Spain}

\author[0000-0001-7925-238X]{Naoki Yoshida}
\affiliation{Department of Physics, The University of Tokyo, 7-3-1 Hongo, Bunkyo, Tokyo 113-0033, Japan}
\affiliation{Kavli Institute for the Physics and Mathematics of the Universe (WPI), UT Institute for Advanced Study, The University of Tokyo, Kashiwa, Chiba 277-8583, Japan}
\affiliation{Research Center for the Early Universe, School of Science, The University of Tokyo, 7-3-1 Hongo, Bunkyo, Tokyo 113-0033, Japan}

\begin{abstract}
Recent JWST observations with superb angular resolution have revealed the existence of clumpy galaxies at high redshift through the detection of rest-frame optical emission lines.
We use the FirstLight simulation to study the properties of (sub-)galactic clumps that are bright in $\OIII 5007\mathrm{\mathring{A}}$ line with flux greater than $\sim 10^{-18} \, {\rm erg\, s^{-1}\, cm^{-2}}$, to be detected by JWST. For 62 simulated galaxies that have stellar masses of $(0.5-6) \times 10^{10} \, M_\odot$ at $z=5$, we find clumps in 1828 snapshots in the redshift range $z = 9.5-5.5$. The clumps are identified by the surface density of star formation rate.
About one-tenth of the snapshots show the existence of clumpy systems with two or more components. Most of the clumps are formed by mergers and can be characterized by their ages; central clumps dominated by stellar populations older than 50 Myr, and off-centered clumps dominated by younger stellar populations with specific star formation rates of $\sim 50 \, {\rm Gyr^{-1}}$. The latter type of young clumps is formed from gas debris in the tidal tails of major mergers with baryonic mass ratios of $1 \leq q < 4$.
The merger-induced clumps are short-lived, and merge within a dynamical time of several tens million years. The number density of the clumpy systems is estimated to be $\sim  10^{-5}\, {\rm cMpc^{-3}}$, which is large enough to be detected in recent JWST surveys.

\end{abstract}


\section{Introduction} \label{sec:intro}
Recent observations by the James Webb Space Telescope (JWST) have significantly advanced our understanding of galaxy formation and evolution in the early Universe. A number of high-$z$ galaxies have been observed so far, and the record of the most distant galaxies has been renewed \citep[e.g.,][]{Curtis-Lake:2023, Wang:2023, Arrabal_Haro:2023, Harikane:2023c, Carniani:2024}. For galaxies at $z \gtrsim 9.5$, strong rest-frame optical emission lines such as \OIII 5007 \AA \, are outside the range of JWST NIRSpec. Additionally, because of the vast distance, these galaxies appear typically to be compact single-component systems \citep[e.g.,][]{Ono:2023, Tacchella:2023}. 
Thus, it is difficult to obtain detailed information on kinematics and internal structure.


Galaxies at $z=6-9$ are promising targets for detailed studies
on clump formation and evolution using emission lines.
NIRCam has a resolution of 0.03"/pixel for long wavelength (2.4-5.0 $\rm \mu \mathrm{m}$), corresponding to $\sim$ 150 pc/pixel at $z=8$ \footnote{NIRCam has a pixel scale of 0.063"/pixel on the sky, but the post-process image data reduction allows the grid resolution 0.03" per pixel \citep{Kashino:2023, Matthee:2023}.}. 
NIRSpec Integral Field Unit (IFU) has a pixel scale of 0.05" \footnote{NIRSpec IFU elements are 0.1"/pixel on the sky, but the combination of dithering and drizzle weighting allows sub-sampling of the detector pixels with 0.05" \citep{Jones:2023, Hashimoto:2023}.}, corresponding to $\sim$ 250 pc/pixel at $z=8$. Gravitational lensing with magnification $\mu$ increases the effective spatial resolution by $\sqrt{\mu}$, allowing detailed studies of internal structure of galaxies \citep[e.g.,][]{Hsiao:2023_MACS0647JD_NIRCam, Bradac:2023, Alvarez-Marquez:2023,Jones:2023, Hashimoto:2023, Morishita:2023, Mestric:2022, Claeyssens:2023, Fujimoto:2024}. 

Especially, recent JWST photometry observations have reported a large number of galaxies at $z > 6$ that contain star-forming clumps and /or elongated structures with a scale of $\sim$ 100 pc in the rest-UV and optical wavelengths. \citep[e.g.,][]{Hainline:2023, Chen:2023, Huertas-Company:2024, Tacchella:2023, Baker:2023}.
A prominent example is JADES-GN-189.18051+62.18047, which consists of five clumps, and four of them are located within a small region of $\sim 7$ kpc. 

Some NIRSpec IFU observations also discovered merging systems of star-forming galaxies and core regions of proto-clusters at $z > 6$ with the strong detection of \OIII 5007 \AA \,\citep[e.g.,][]{Hashimoto:2023, Jones:2023, Arribas:2023, Venturi:2024}. \citet{Hashimoto:2023} and \citet{Jones:2023} found over four components within a $\sim$ 11 kpc regions. 
While these observations suggest a rapid assembly of high redshift galaxies, it remains largely unknown how the compact clumpy systems emerge in the early epochs.

Clump formation at lower redshift $z = 0-4$ have been widely investigated both by observations \citep{Genzel:2011, Tadaki:2014, Guo:2015, Elmegreen:2013} and theoretical studies \citep{Oklopcic:2017, Ceverino:2010, Bournaud:2009, Agertz:2009, Romeo:2010, Romeo:2014}. They have proposed two major mechanisms for clump formation: violent disk instability (VDI) and mergers.

The former is driven by, for example, cold gas accretion flowing along the cosmic web \citep{Dekel:2009, Ceverino:2010, Dekel:2013, Rodriguez-Gomez:2016}, and the local disk instability is regulated by the Toomre parameter $Q$ \citep{Toomre:1964}. This is originally defined as $Q \equiv \sigma_{\rm r} \kappa/\pi G \Sigma$ for an infinitesimally thin gas disk, where $\sigma_{\rm r}, \kappa, G, \Sigma$ are the radial velocity dispersion, the epicyclic frequency, the gravitational constant, and gas surface density. Clumps formed in massive disk galaxies have been studied by cosmological simulations \citep{Agertz:2009, Ceverino:2010, Mandelker:2014, Mandelker:2017, Buck:2017, Ceverino:2023, Inoue:2016, Genel:2012} as well as isolated disk models \citep{Tamburello:2017, Faure:2021, Bournaud:2014, Bournaud:2009, Hopkins:2012}. Observations also identified clumps induced by VDI with significant rotations \citep[e.g.,][]{Genzel:2006, Genzel:2008}. The fate of formed clumps (survival and disruption) has often been discussed from theory and observations \citep{Dekel:2009, Renaud:2014, Elmegreen:2007}.

The later competitive mechanism is mergers. Galaxy mergers cause significant gas density perturbations and even promote star formation \citep[e.g.,][]{Di_Matteo:2008}, and thus may be another promising formation path for clumps. 
Such clumpy galaxies via mergers are often discussed with a major-merger rate or a pair fraction. Numerical simulations suggest that the galaxy merger rate may increase at higher redshifts \citep[e.g.,][]{Genel:2008, Genel:2009, Fakhouri:2009, Hopkins:2010, Rodriguez-Gomez:2015}. 

Since observations cannot measure the rate directly, they instead quantify pair fractions \citep[e.g.,][]{Tasca:2014, Ribeiro:2017, Lopez-Sanjuan:2013}, which has been statistically investigated at lower redshift $z \lesssim 6$ with a large samples by HST. \citet{Ribeiro:2017} argued that the obtained pair fraction infers a major merger fraction of $\sim$20 \% over the redshift range $2 <z < 6$.

In addition to a pair fraction, a clumpy fraction $f_{\rm clumpy}$ has also been investigated to evaluate the contribution of merger and VDI \citep[e.g.,][]{Guo:2015, Shibuya:2016, Ribeiro:2017}. \citet{Ribeiro:2017} found that a clumpy fraction increases from 35 to 50 percent from z = 2-6, while \citet{Shibuya:2016} reported the opposite trend and suggested that VDI is the main origin of clumps. The clumpy fraction may also depend on stellar mass \citep{Guo:2015}, making it difficult to understand the redshift evolution.


While clumpy galaxies at $z > 7$ are newly observed by JWST in multiwavelength and their stellar mass and size are statistically investigated \citep[e.g.,][]{Mestric:2022, Claeyssens:2023}, the redshift evolution of clumpy fractions and constraints on the formation scenario are not well understood. For numerical simulations, large-scale cosmological simulations such as Illustris TNG \citep{Pillepich:2018}, THESAN \citep{Shen:2024}, FLARES \citep{Roper:2022}, BLUETIDES \citep{Marshall:2022}, and SIMBA \citep{Wu:2020} have a minimum resolution of a few hundred pc, making it hard to resolve internal structures of galaxies at high redshift.

Zoom-in simulations can achieve better resolution, and clumpy structures are also reported \citep{Barrow:2017, Ma:2018, Katz:2019, Arata:2020, Gelli:2020, Gelli:2021, Kohandel:2020, Kimm:2016, Calura:2022}. However, their samples are often limited to only a few galaxies, and there are few theoretical studies at $z > 6$ that qualitatively and statistically discuss the formation of clumps and their fates, their physical properties, and the clumpy fractions.

In this paper, we use the FirstLight simulation suite, which achieves a very high resolution of $17$ pc and provides a large sample size to allow robust statistical analysis. We use 
outputs for 62 simulated galaxies with 4092 snapshots in $z =6-9$. Our main objective is to study the properties of galaxy clumps at the epoch of reionization, and their formation and evolution.

In our previous study \citep{Nakazato:2023}, we devised a physical model of \HII regions and calculated the luminosity of emission lines for the same FirstLight galaxy samples. The resulting relation between the star formation rate and \OIII luminosity is consistent with recent observations \citep[e.g.,][]{Hashimoto:2018, Tamura:2019, Harikane:2020}. We use the same model to characterize clumpy galaxies that are bright in rest-frame optical emission lines such as \OIII 5007 \AA. 


The rest of the present paper is organized as follows. In Section \ref{sec:method}, we introduce the numerical simulation
and describe the method to calculate line luminosities
and define clumps. In Section \ref{sec:result}, we discuss in detail the formation and evolution of a few clumpy systems.
Summary and conclusions are given in Section \ref{sec:discussion}. Throughout this paper, we assume a set of parameters in $\Lambda$CDM cosmology with $\Omega_{\rm m} = 0.307, \, \Omega_{\rm b} = 0.048, \, h = 0.678$ and $\sigma_8 = 0.823$ from \textit{Planck13} results \citep{Planck:2014}.
\section{Method} \label{sec:method}

\subsection{Zoom-in Simulation} \label{subsec:simulation}
We use the cosmological zoom-in simulation suite FirstLight \citep{Ceverino:2017}. The simulations are performed using the \textsc{art} code \citep{Kravtsov:1997, Kravtsov:2003, Ceverino:2009}, which follows gravitational $N$-body dynamics and Eulerian hydrodynamics with Adaptive Mesh Refinement (AMR). The parent simulation has a box size of 40 comoving $h^{-1}\, \mathrm{Mpc}$ on a side. We identified the 62 most massive halos in the parent dark matter-only simulation. The halos have maximum rotation velocities with $V_{\rm max} > 178~\mathrm{km}\, {\rm s}^{-1}$ at $z = 5$.
We then performed zoom-in simulations in these regions with an effective resolution of $4096^3$ and a maximum resolution of 17-32 proper pc. The dark matter particle mass is $m_{\rm DM} = 8\times 10^4\, M_\odot$ and the minimum stellar particle mass is $10^3\, M_\odot$. \notice{We have compared the stellar-to-halo mass relation with other simulations with the same setup but a higher resolution of $m_{\rm DM} = 10^4\, M_\odot$ and confirmed that the convergence is satisfactory \citep{Ceverino:2023, Ceverino:2024}.} A number of baryonic processes in star formation and feedback are incorporated in a sub-grid manner. We describe them briefly as follows. 

The zoom-in simulations follow radiative cooling by atomic hydrogen and helium, metal ions and atoms, and molecular hydrogen, and photoionization heating by a time-dependent UV background with partial self-shielding \citep{Haardt:1996}.

Stellar particles are created in gas cells according to the criteria for the hydrogen number density of $n_{\rm H} > 1\, {\rm cm^{-3}}$ and temperature of $T < 10^4\, {\rm K}$. We use a stochastic model with the probability that scales with the gas free-fall time \citep{Ceverino:2009}. Our simulations incorporate a constant heating rate for 40 Myr after star formation as thermal stellar feedback driven by stellar winds and supernovae (SNe). The code also implements radiative feedback from massive young stars as a non-thermal pressure term \citep{Ceverino:2014}. 
The kinetic feedback model also includes the injection of momentum coming from the (unresolved) expansion of gaseous shells from supernovae and stellar winds \citep{Ostriker:2011}.  More details can be found in \cite{Ceverino:2017}, \cite{Ceverino:2014}, \cite{Ceverino:2010}, and \cite{Ceverino:2009}.

Our simulation implements a pressure floor to prevent artificial fragmentation \citep{Truelove:1997, Machacek:2001, Robertson:2008, Agertz:2009, Rosdahl:2015}. \notice{In order to resolve Jeans length at least seven cells ($N_{\rm J} = 7$), we set the artificial pressure floor at 
\begin{equation}
    P_{\rm floor} = \frac{G\rho_{\rm gas}^2 N_{\rm J}^2 \Delta x_{\rm min}^2}{\pi \gamma}, 
\end{equation}
where $\Delta x_{\rm min}$ and $\gamma = 5/3$ are the minimum cell size and the specific heat ratio for monoatomic gas. The pressure in the Euler equation is replaced by an effective pressure that is set to $P_{\rm floor}$ when it would have a lower value. The threshold number of cells (i.e., $N_{\rm J}=7$) is beyond the standard criteria \citep[$N_{\rm J}=4$, ][]{Truelove:1997} and this shows good convergence in cosmological simulations of clumpy galaxies \citep{Ceverino:2010}. Note that this implementation prevents artificial fragmentation but it also prevents the growth of structures below $\sim$ 100 pc scales. Previous simulations with the same implementation and resolution have already investigated low-redshift clumps \citep{Mandelker:2014, Mandelker:2017, Inoue:2016, Inoue:2019, Ceverino:2023}. More details are provided in Appendix \ref{subsec:artificial_pressure_floor}.}


Our 62 galaxy samples have stellar masses greater than $\sim 10^{10} M_{\odot}$ at $z=5$. The maximum resolution in the zoom-in hydro simulations, particularly valid in dense clumps, enables the simulations to resolve gas densities of $\sim 10^3 \,{\rm cm^{-3}}$ with temperatures of $\sim 300\, {\rm K}$. 
We store a total of 66 snapshots between $z=9.5$ to $z = 5.5$ for each galaxy, with a spacing of the cosmic expansion parameter $\Delta a = 0.001$, corresponding to $7-10$ Myr \footnote{The time step width in the simulation is much shorter than this, being typically 1000 years.}. The frequent output timing is sufficient to follow dynamics during galaxy mergers that occur on the tidal timescale of 10-100 Myrs \citep{Renaud:2013}.
\subsection{Line luminosity Calculation} \label{subsec:line_luminosity_calculation}
We calculate the luminosities of emission lines from \HII regions for the simulated galaxies and clumps. We adopt essentially the same model as \citet{Nakazato:2023} and assume that each stellar particle is surrounded by a spherical \HII region. 
We generate a table of a set of emission lines using \textsc{cloudy} \citep{Ferland:2013}. As an input stellar SEDs, we adopt a single star model of BPASS \citep{Eldridge:2017}. According to \citet{Eldridge:2017}, these SEDs are similar to other ones such as Starburst 99 \citep{Leitherer:1999}, GALAXEV \citep{Bruzual_Charlot:2003}, and \citet{Maraston:2005} at the age of $<$ 1 Gyrs. Note that BPASS binary star model can enhance the number of ionizing photons by a factor of up to $\sim$ 10 at the age of $<$ 10 Myr, leading to the [O{\sc iii}] flux increase up to $\sim$ 1.2 times higher than that of the single model \citep{Ceverino:2021}.
The input parameters for \textsc{cloudy} are gas metallicity $Z_{\rm gas}$, ionization parameter $U$, and gas density in \HII regions $n_{\textsc{hii}}$. The adopted values of the parameters are listed in Table \ref{table:cloudy}. 
In our table, we normalize the emission line luminosity for each stellar particle $(L_{{\rm line}, i})$ by H$\beta$ luminosity as 
\begin{equation}
    L_{{\rm line}, i} = (1 - f_{\rm esc}) C_{\rm line}(Z_{\rm gas}, U, n_{\textsc{hii}}) L^{\rm caseB}_{{\rm H}\beta}.
\end{equation}
Here $f_{\rm esc}$ is the escape fraction and we assume $f_{\rm esc} = 0.1$, a value suggested by both radiation hydrodynamics simulations \citep{Yajima:2011, Kimm_Cen:2014, Wise:2014, Trebitsch:2017, Xu:2016}, and observations \citep[e.g.,][]{Nakajima:2020} for high-$z$ galaxies. Since individual \HII regions are not resolved in our simulations, we assume that the gas density of a \HII region is given by 
\begin{equation}
    n_{\textsc{hii}} = \left\{
    \begin{array}{ll}
    100 \, {\rm cm^{-3}}& (n_{\rm gas, grid} < 100\, {\rm cm^{-3}})\\
    n_{\rm gas, grid} & (n_{\rm gas, grid} \geq 100\, {\rm cm^{-3}}),
    \end{array}
    \right. .
\end{equation}
where $n_{\rm gas, grid}$ is the gas number density of a grid with a size of $\Delta = 50\, {\rm pc}$. Our motivation for using a variable \HII region density comes from the need to account for \notice{grid-to-grid density variations} in our simulations. Setting a fixed \HII region density at a typical value of $100 \, {\rm cm^{-3}}$ \citep{Osterbrock:2006, Hirschmann:2017, Hirschmann:2022} could lead to unrealistic conditions where the grid density exceeds this value. To address this, we adopt a model with a variable \HII region density if $n_{\rm gas, grid} > 100 \, {\rm cm^{-3}}$\footnote{\citet{Nakazato:2023} assumed the fixed \HII density as $n_{\textsc{hii}} = 100\, {\rm cm^{-3}}$. 
We also have checked the dependency of \OIII 5007 \AA \, luminosity on the density model and found that the value changes within only $\lesssim$ 1 \%.}. Our model results in a higher global-averaged \HII region density in high-$z$ $(z > 6)$ simulated galaxies compared to typical local \HII regions. This trend is also supported by recent JWST observations \citep[e.g.,][]{Isobe:2023, Fujimoto:2022_z8p5, Abdurro'uf:2024}.


We look for the line ratio table $C_{\rm line}$ with the three parameters $Z_{\rm gas}, \, U, \, n_{\textsc{hii}}$ which are closest to those of the target stellar particles in our simulation.
We obtain the line luminosity for each grid as $L_{\rm line, grid} = \sum^N_i L_{{\rm line},i}$, where $N$ is the number of stellar particles that the grid contains. 

\begin{table}[htbp]
\centering
\begin{tabular}{lr}
\hline
\vspace{0.4mm}
$\log_{10}~ (Z_\mathrm{gas}/Z_\odot)$ & -1.30, -0.70, -0.40, 0., 0.30   \\
$\log_{10}~ U$           & -4.0, -3.9, ..., -1.1, -1.0 \\
$\log_{10}~ (n_\textsc{Hii}/\mathrm{cm}^{-3}$) &  1.0, 2.0, 3.0 \\
\hline
\end{tabular}
\caption{The parameters used to calculate the line luminosities with \textsc{cloudy}.}\label{table:cloudy}
\end{table}

In our previous study \citep{Nakazato:2023, Mushtaq:2023}, we find that the magnitude of dust attenuation at rest-frame 5007 \AA \, is approximately $A_{5007} = 0.08, 0.24, 0.7$ for individual {\it galaxies} with $M_*/M_\odot = 10^8, 10^9, 10^{10}$, respectively. Thus, we expect that dust attenuation affects the line luminosity of a small clump by a factor of a few. We further discuss this issue of dust attenuation in Section \ref{sec:discussion}.

\subsection{Clump Finder} \label{subsec:clump_finder}
In order to compare the properties of the simulated clumpy galaxies with observations, we identify clumps in 
two-dimensional ``images" using projection along a random line of sight ($z$-axis) of a cube with a fixed side length of 10 kpc. The size is larger than the main central galaxy
, and is close to a field of view of NIRSpec IFU at $z=8$. 
We assign masses of gas, stars, and dark matter to a uniform grid with $\Delta = 50\, {\rm pc}$ by using a cloud-in-cell interpolation also employed in \citet{Mandelker:2014,Mandelker:2017} and \citet{Ceverino:2023}. The value is \notice{about three} times the maximum AMR resolution.
We first mark grids that contain young ($<$ 10 Myr) stars with their surface mass density ($\Sigma_{M_{*} {\rm (young)}}$) greater than $10^{8.5} \, M_\odot{\rm kpc^{-2}}$, corresponding to the surface SFR density of $10^{1.5} M_\odot{\rm yr^{-1} kpc^{-2}}$. We then identify candidate groups by running a Friends-of-Friends finder with linking length $\Delta$ to the marked grids. 
Groups with at least 16 grids ($N_{\rm grid}$) are identified as clumps. The threshold grouping number of grids corresponds to the minimum radius of clump $R_{\rm min} = \sqrt{N_{\rm grid} \Delta^2/\pi} = 113 \, {\rm pc}$.
The threshold surface stellar mass density and the minimum number of grids are chosen to ensure that the identified clumps are resolved by observations by NIRSpec IFU and NIRCam with the line flux of $F_{\OIII} \gtrsim 10^{-18} \,{\rm erg\, s^{-1} \, cm^{-2}}$ at $z = 8$.
The details of threshold determination are described in the next section \ref{subsec:parameter_determination}. 
 We have also checked the effect of line of sight by changing inclination, and it does not change the number of clumps.

\begin{figure}
    \centering
    \includegraphics[width = \linewidth, clip]{\figdir/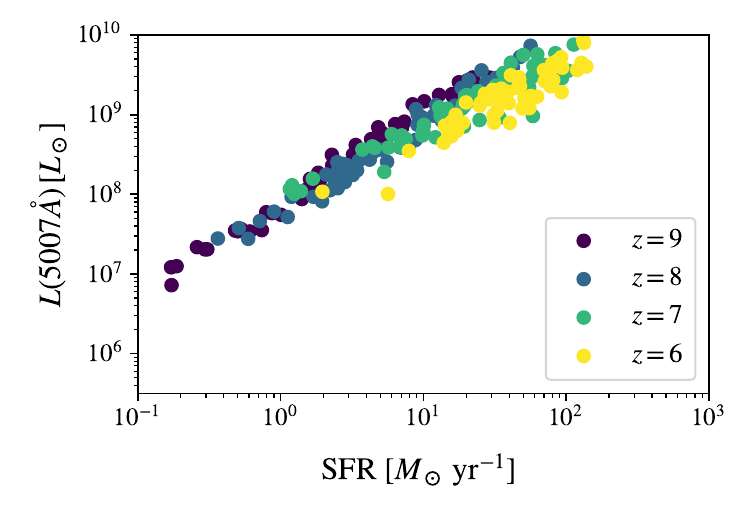}
    \caption{ The \OIII 5007 \AA \, luminosity vs. SFR for our 62 simulated galaxies at $z=9$ (purple), 8 (blue), 7 (green), and 6 (yellow). 
}
    \label{fig:LOIII-SFR}
\end{figure}
\subsubsection{Parameter determination}\label{subsec:parameter_determination}
We set three parameters for our clump finder; grid size ($\Delta$), threshold surface mass density of young stellar population ($\Sigma_{M_{*} {\rm (young)}}$), and the minimum number of grids ($N_{\rm grid}$).  
We assume that a \HII region exists around each stellar particle. 
The grid size is set to be larger than the typical Str\"{o}mgren radius for each stellar particle in our line emission model.
The Str\"{o}mgren radius $R_{\rm s}$ is given by 
\begin{align}
    R_{\rm s} &= \left(\frac{3 Q_0}{4\pi n^2_{\textsc{HII}}\alpha_{\rm B}}\right)^{1/3} \\
              &= 5.4 \, {\rm pc} \left(\frac{n_{\textsc{HII}}}{10^2\, {\rm cm^{-3}}}\right)^{-2/3} \left(\frac{Q_0}{5\times 10^{49}\, {\rm s^{-1}}} \right)^{1/3} , 
\end{align}
where $n_{\rm HII}$ is the gas density in \HII regions and $Q_0$ is the number of ionizing photons per unit time. 
The electron temperature is $T_{\rm e} = 10^4\, \mathrm{K}$, and a filling factor of $\epsilon = 1$ is assumed. The value of $Q_0$ depends on stellar age, mass, and spectrum, which means the value varies to simulation resolutions and model SEDs. We have checked the distribution of $Q_0$ values for all stellar particles in our simulations and obtained the median (maximum) value of $Q_0 = 5 \times 10^{50}\, (5\times 10^{52}) \,{\rm s^{-1}}$, and the corresponding Str\"{o}mgren radius is $R_{\rm s} \sim 15 \, (54) \, {\rm pc}$ for $n_{\rm HII} = 100\, {\rm cm^{-3}}$. We thus set $\Delta = 50\, {\rm pc}$. 

To determine the threshold of surface young stellar mass density ($\Sigma_{M_{*} {\rm (young)}}$), we consider the relationship between $L_{\OIII 5007\mathrm{\mathring{A}}}$ and 
star formation rate (SFR). This relationship is implied from pre-JWST observations that have shown the tight correlation between [O{\sc iii}] 88 \mum and SFR at $z > 6$ \citep[e.g.,][]{Harikane:2020, Moriwaki:2018, Katz:2019, Pallottini:2022}.  There are still few statistical investigations of [O{\sc iii}] 5007 \AA \, emitters at $z > 6$ \citep{Matthee:2023, Meyer:2024}, but lower-redshift $(z \lesssim 2)$ galaxies are already reported to have tight correlation between [O{\sc iii}] 5007 \AA \, and SFR by both observations and simulations \citep[e.g.,][]{Villia-Velez:2021, Hirschmann:2023}.
Figure \ref{fig:LOIII-SFR} shows the \OIII 5007 \AA \,luminosity against SFR for our galaxy samples.
It is approximated as\footnote{
Our linear fitted line for each redshift is
\begin{align*}
\log_{10} L_{\OIII}  [L_\odot] &= 1.16 \times \log_{10} {\rm SFR} + 7.89  ~~(z=9) \\
\log_{10} L_{\OIII}  [L_\odot] &= 1.16 \times \log_{10} {\rm SFR} + 7.74  ~~(z=8) \\
\log_{10} L_{\OIII}  [L_\odot] &= 0.87 \times \log_{10} {\rm SFR} + 7.97  ~~(z=7) \\
\log_{10} L_{\OIII}  [L_\odot] &= 0.97 \times \log_{10} {\rm SFR} + 7.67  ~~(z=6) .
\end{align*}
}
\begin{align}
  L_{\OIII 5007\mathrm{\mathring{A}}} &\sim 6\times 10^7 \left(\frac{{\rm SFR}}{M_\odot {\rm yr}^{-1}}\right) L_\odot  \\
  &\sim 6 \times \left(\frac{M_{* {\rm (young)}}}{M_\odot}\right) \, L_\odot, \label{eq:LOIII_SFR}
\end{align}
where $M_{* {\rm (young)}}$ is the mass of very young ($<$ 10 Myrs) stars, and the instantaneous SFR is estimated as $(M_{* {\rm (young)}}/ 10^7)\, M_\odot\,{\rm yr}^{-1}$.
The flux limit to detect a clump by \OIII 5007\AA\, line is $\sim 3\times 10^{-18} \, {\rm erg\, s^{-1}\, cm^{-2}}$, which corresponds to signal-to-noise ratio of $S/N>5$ with an exposure time 
 of $\sim 10^4$ sec \citep{Hashimoto:2018, Jones:2023, Arribas:2023, Matthee:2023}. 
The corresponding luminosity is $L_{\OIII 5007\mathrm{\mathring{A}}} \sim 6\times 10^7 \, L_\odot$ at $z=8$. Substituting this into eq.(\ref{eq:LOIII_SFR}), we obtain $M_{*, {\rm (young)}} \sim 10^7\, M_\odot$ for a ``detectable" clump. 

We set the minimum radius to be $R_{\rm clump}\sim 100$ pc, which can be resolved by recent JWST NIRCam/NIRSpec IFU observations \citep[e.g.][]{Chen:2023, Hainline:2023, Hashimoto:2023, Harikane:2024}. We therefore set the threshold grid grouping number as $N_{\rm grid} = 16$, corresponding to $R_{\rm clump} = \sqrt{N_{\rm grid} \Delta^2/\pi}= 113 \, {\rm pc}$, where one grid area is $\Delta^2 = 50^2 \,{\rm pc^2}$. The minimum clump size is around seven times larger than the minimum AMR resolution of 17 pc.
\notice{We emphasize that our identified clumps correspond to unlensed observed clumps. Strong lensing can enhance the resolution up to the $\sim$ 10 pc level \citep[e.g.][]{Adamo:2024, Bradley:2024, Mowla:2024}.}

For this minimum clumps to have ``detectable” luminosity, i.e., to have young stellar mass of $M_{*({\rm young})} \gtrsim 10^7 \, M_\odot$, each grid should have a surface stellar mass density of $\Sigma_{M_{*({\rm young})}} \geq 10^7/16 \, M_\odot{\rm grid^{-1}} = 10^{8.5} \, M_\odot{\rm kpc^{-2}}$. This value also corresponds to the surface SFR density of $\Sigma ({\rm SFR}) \geq 1/16 \, M_\odot {\rm yr^{-1}\, grid^{-1}}= 10^{1.5} \, M_\odot{\rm  yr^{-1}kpc^{-2}}.$


\begin{figure*}
    \centering
    \includegraphics[width = \linewidth, trim = 1 0 0 1, clip]{\figdir/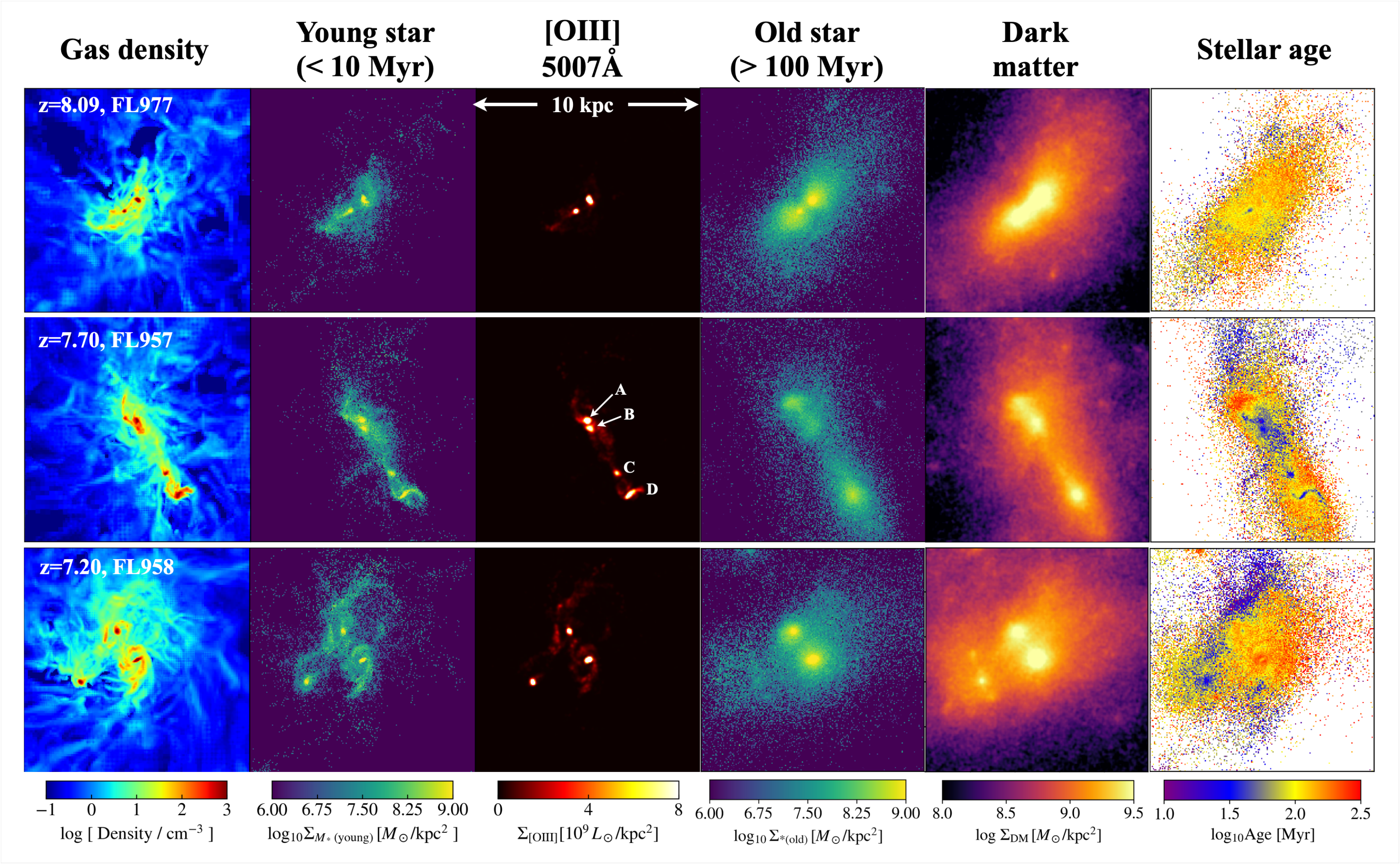}
    \caption{Three simulated clumpy galaxies at $z=9.00$ (top, ID=FL977), $z=7.70$ (middle, ID=FL957), and $z=7.20$ (bottom, ID=FL958). Each column from left to right represents the averaged gas density, surface stellar mass density of young ($<$ 10 Myr) population, \OIII 5007 \AA \,line flux, surface mass density of old ($>$ 100 Myr) stellar population and dark matter, and mass-weighted stellar age, respectively. 
    Each panel has a side length of 10 proper kpc and the same size of projected depth.}
    \label{fig:projection}
\end{figure*}
\section{Result} \label{sec:result}
\subsection{Projection of physical properties for simulated clumpy galaxies}
Figure \ref{fig:projection} shows three examples of simulated galaxies. Two, four, and three clumps are identified in these outputs as bright components. The clumps trace the distribution of young stars but also that of dense gas with $n_{\rm gas} \gtrsim 100\, {\rm cm^{-3}}$. Tail features can also be seen in \OIII emission, similar to the recent galaxies observed by JWST such as MACS1149-JD1 \citep[$z = 9.1$, ][]{Bradac:2023} and B14-65666 \citep[$z=7.15$, ][]{Sugahara:2024}. The distribution of the stars older than 100 Myrs appears more diffuse than the young stellar population. 
The underlying dark matter distribution shows signatures of mergers, which will be explained in detail in Section \ref{subsec:clump_formation}. Some clumps (e.g., clump C in Figure \ref{fig:projection}) are located off-center. The clumps are separated with $\sim 0.1-4 \, {\rm kpc}$ distances, similarly to the clumps at $z\simeq 6-8$ 
studied by \citet{Chen:2023}, with typical separations of $\sim $ 0.3 - 4.3 kpc. Such close clumps are also found in other JWST observations at $z>6$; the JADES \citep{Hainline:2023}, EIGER \citep{Matthee:2023} and CANUCS \citep{Asada:2024} surveys.
The stellar age distributions in the right-most panels 
show a clear difference in the distribution of young and old components. Interestingly, similar features are found by JWST MIRI observations \citep[e.g.,][]{Colina:2023}.

The panels in the middle row focus on the galaxy FL957, which contains the highest number of clumps at $z > 6$. 
To see if those clumps are gravitationally bound or not, we have estimated the virial parameter \citep{Bertoldi:1992}, $\alpha_{\rm vir} \equiv 5 \sigma^2_{\rm clump} R_{\rm c}/G M$, where $\sigma_{\rm clump}$ and $ R_{\rm c}$ are velocity dispersion within a clump and clump radius. All clumps have $\alpha \sim 0.2$, which indicates that these clumps are gravitationally bound.

The four-clump feature motivates us to compare it with the recent observation by \citet{Hashimoto:2023}, who spectroscopically studied the core of the most distant protocluster at $z=7.88$, A2744-z7p9OD, with JWST NIRSpec integral field spectroscopy. The core region contains four clumps which are detected by \OIII 5007 \AA \, in a region of $\sim 11 \times 11 \, {\rm kpc}$. 
Table \ref{table:FL957_a0p115} summarizes the clump properties. We find that the stellar mass, SFR, and \OIII luminosities are similar to the observed clumps (see Table 1 in \citet{Hashimoto:2023}). 
These properties are typical for clumps younger than 50 Myrs identified in the multiple systems, as shown in Table \ref{table:clump_properties}, which is discussed in the next section. We discuss that such properties are different from clumps older than 50 Myr in Section \ref{subsec:clump_properties}. In Sections \ref{subsec:clump_formation} and \ref{subsec:clump_fate}, we show the formation and evolution of multi-clump systems. 

\begin{table*}[]
\centering
\begin{tabular}{ccccccc}
\hline 
\multicolumn{1}{l}{} & $M_{*, {\rm (young)}}$  $[M_\odot]$ & $M_{*, {\rm  (all)}}$ $[M_\odot]$ & SFR $[M_\odot/{\rm yr}]$  & $L_{\OIII 5007\mathrm{\mathring{A}}}\, [L_\odot]$  & $R_{\rm c}\, [{\rm pc}]$  & Age [Myr]\\ \hline\hline
A                    & $8.32\times 10^7$   & $1.32\times 10^8$ & 8.32 & $1.10\times 10^9$  & 157   &  39\\
B                    & $4.62\times 10^7$   & $1.87\times 10^8$ & 4.62 & $5.59\times 10^8$  & 138  &  27 \\
C                    & $2.45\times 10^7$   & $3.76\times 10^7$ & 2.45 & $2.85\times 10^8$  & 116  & 34 \\
D                    & $8.49\times 10^7$   & $3.32\times 10^8$ & 8.49 & $9.14\times 10^8$  & 155  & 30 \\\hline
\end{tabular}
\caption{Summary of properties of FL957 at $z=7.70$, which contains four clumps within 10 $\times$ 10 kpc region. Note that $M_{*, ({\rm young})}$ refers to stellar masses younger than 10 Myrs and $M_{*, ({\rm all})}$ are stellar masses for all ages. A clump radius $R_{\rm c}$ in the 5th column is obtained as $R_{\rm c} = \sqrt{\Delta^2 N_{\rm grid}/\pi}$.}
\label{table:FL957_a0p115}
\end{table*}

\begin{table*}[]
\centering
\begin{tabular}{l|ccccc}
\hline
    & \# of clumps &$M_{*, {\rm (all)}}$  $[M_\odot]$ & $n_{\rm gas, clump}\, [{\rm cm^{-3}}]$ & Age [Myr] & $R_{\rm c}\, [{\rm pc}]$ \\ \hline\hline
single system                 & 1503    & $5.2\times 10^9$    & 283    & 134    & 234          \\
clumpy system (total)         & 751     & $2.5\times 10^9$    & 241    & 108    & 180          \\
clumpy system ($<$ 50 Myr)    & 182     & $2.7\times 10^8$    & 269    & 35     & 142          \\
clumpy system ($>$ 50 Myr)    & 569     & $3.2\times 10^9$    & 232    & 131    & 192          \\
(single $+$ clumpy) system    & 2254    & $4.3\times 10^9$    & 269    & 125    & 216        
\\\hline
\end{tabular}
\caption{Summary of physical properties of galaxies we identify in single systems and clumpy systems. The clumps in clumpy systems are further categorized into two groups: those with ages younger than 50 Myr and older than 50 Myr. The last row of the table lists the average values for both systems. We see that old clumps ($>$ 50 Myr) in clumpy systems have similar properties to ones in single systems, as discussed in section \ref{subsec:clump_properties}.}
\label{table:clump_properties}
\end{table*}

\subsection{Distributions of clump properties}\label{subsec:clump_properties}
\begin{figure}[ht]
    \centering
    \includegraphics[width = \linewidth, clip]{\figdir/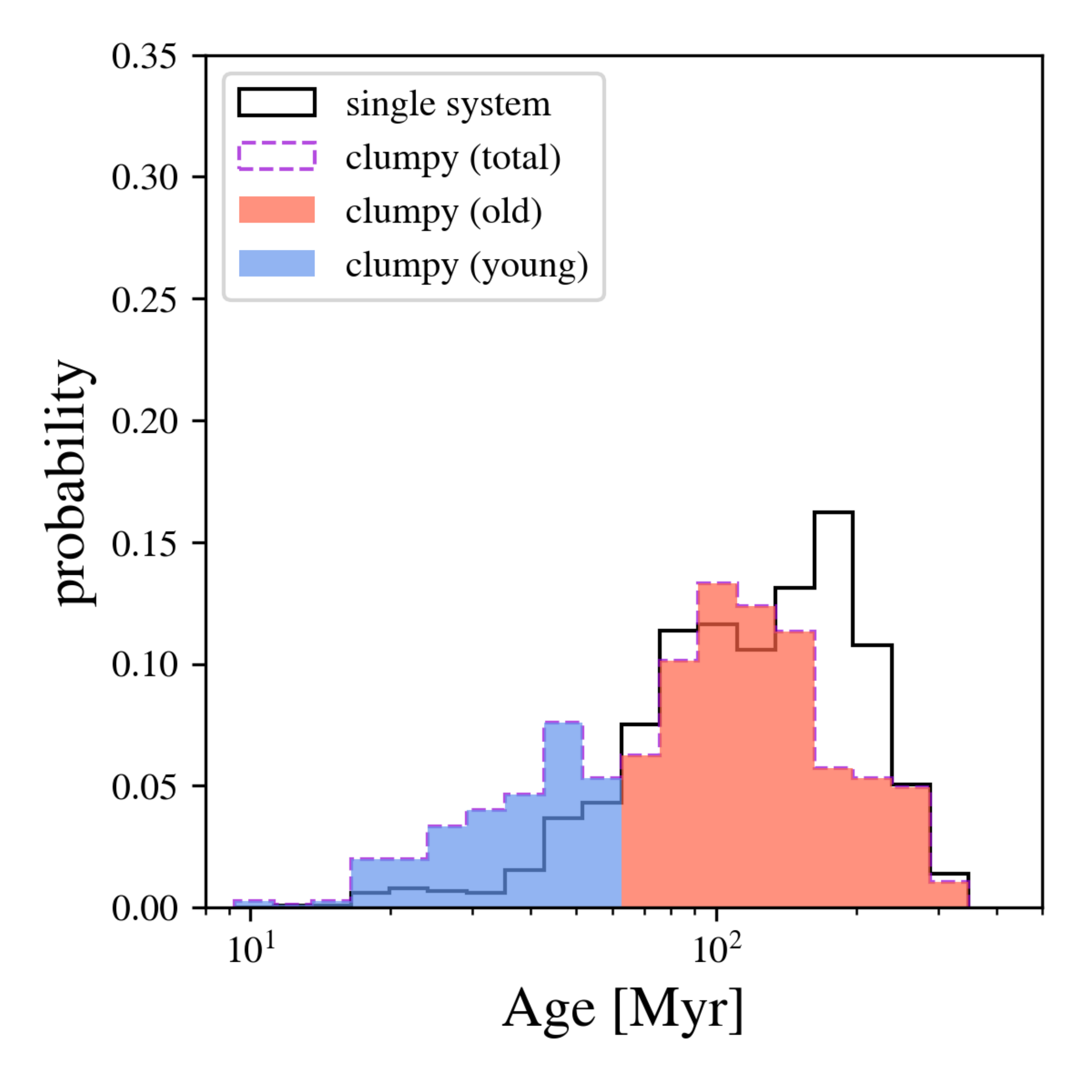}
    \caption{The distribution of mass-weighted clump ages. Black and dashed-purple histograms represent single systems (detected one central clump, i.e., proto-bulge) and clumpy systems, respectively. The histograms are normalized by the total number of clumps for each system. We divide the clumpy system into clumps younger than 50 Myrs (blue) and those older than 50 Myrs (red). }    \label{fig:age_clump_normalized}
\end{figure}

Our clump finder identifies 2254 clumps in total. Of these, 1503 clumps are single systems, i.e., galaxies with one clump, often appearing as proto-bulge components. The remaining 751 clumps are in 325 clumpy systems. Table \ref{table:clump_properties} summarizes the physical properties of the clumps. 
Figure \ref{fig:age_clump_normalized} shows the distribution of the clump age weighted by stellar mass for single and multi-clump objects. The distribution for single systems has a peak at $\sim 200$ Myr, while the clumps in clumpy systems have a peak at $\sim 90$ Myr. 
Based on the age distribution, 
we can separate them into two types of clumps: clumps younger than 50 Myr and clumps older than 50 Myr, which are indicated by blue and red histograms, respectively.
\begin{figure*}[ht]
    \centering
    \includegraphics[width = \linewidth, trim = 1 2 0 1, clip]{\figdir/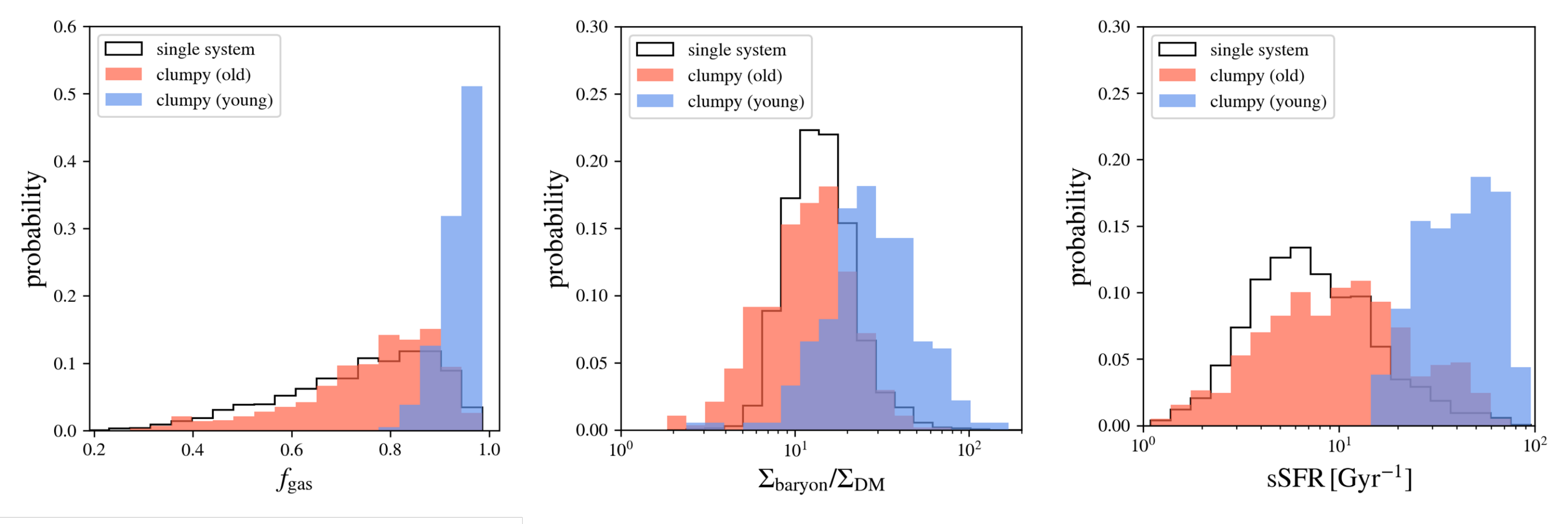}
    \caption{The probability distribution of the gas fraction, baryon-to-dark matter mass ratio, and specific star formation rate for our clump samples. The histograms with black solid lines are for single systems, and we color the histograms for clumps depending on their age. The distributions are normalized to the total number of clumps in each of the colored types.}
    \label{fig:clump_property_statistics}
\end{figure*}
Table \ref{table:clump_properties} also shows the properties of clumps younger than 50 Myr and older than 50 Myr, respectively. We see that old clumps have similar stellar mass, age, and clump size to the clumps identified in single systems. Young clumps are smaller and their stellar mass is ten times lower than older clumps.
Figure \ref{fig:clump_property_statistics} shows the distributions of three properties for different systems: single systems with one clump ($N_{\rm clump} =1$, black solid line), and young and old clumps in clumpy systems
($N_{\rm clump} \geqq 2$, red and blue). The probability distributions 
are normalized to the total number of clumps in each of the three systems. 


The left panel of Figure \ref{fig:clump_property_statistics} refers to the gas fraction ($f_{\rm gas} \equiv M_{\rm gas}/M_{\rm baryon}$) of individual clumps. Single systems have a distribution in the range of $f_{\rm gas} \sim 0.4 - 0.95$ and have the peak at roughly $f_{\rm gas} \sim 0.85$. While clumps younger than 50 Myrs concentrate on the value of 0.95, clumps older than 50 Myrs have almost the same distribution as single systems. 

The middle panel of Figure \ref{fig:clump_property_statistics} refers to the surface mass ratio of baryon and dark matter. Single systems have a peak at $\Sigma_{\rm baryon}/\Sigma_{\rm DM} \sim 10$, and the clumps older than 50 Myr have almost the same distribution and peak value. The shape of the distribution of clumps younger than 50 Myr is similar to the old ones, but the peak is shifted to a 0.5 dex larger value. This implies that over 50\% of young clumps are formed off-center of their host dark matter halos, i.e., in a baryonic-rich environment.

The right panel of Figure \ref{fig:clump_property_statistics} refers to the sSFR of individual clumps. The peak of the distribution for single systems is at sSFR $\sim 6\, \mathrm{Gyr^{-1}}$, while the peaks for old and young clumps are at $\sim 10$ and $ 55 \, \mathrm{Gyr^{-1}}$, respectively. The sSFR values of single systems and old ($> 50 {\rm Myr}$) clumps in clumpy systems are similar to that of main-sequence galaxies at $z \sim 6$, ${\rm sSFR} \sim 6 \, {\rm Gyr^{-1}}$ \citep{Popesso:2023}. However, young clumps in clumpy systems have over 9 times larger sSFR than that of main-sequence galaxies, and such large values are also seen in observed merging galaxies at $z> 6$ \citep[e.g.,][]{Sugahara:2024}.

From the three physical properties in Figure \ref{fig:clump_property_statistics}, we see that clumps older than 50 Myr tend to have the same properties as single systems. On the other hand, clumps younger than 50 Myr are located in baryonic-rich environments. They are also gas-rich and in the process of bursty star formation. This indicates a different formation path for young clumps.


\begin{figure*}[ht]
    \centering
    \includegraphics[width = \linewidth, clip]{\figdir/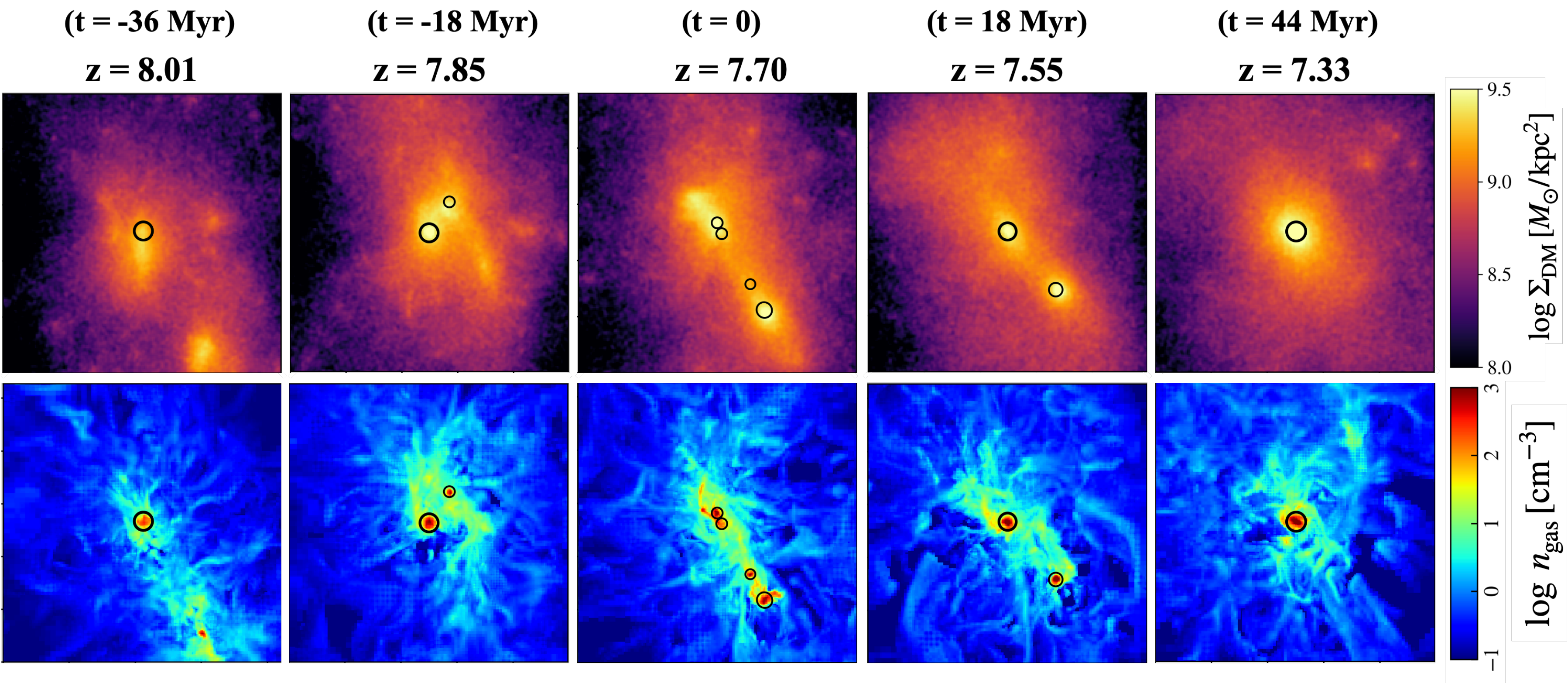}
    \caption{ Time evolution of dark matter density (top) and gas density distribution (bottom) from $z = 8.01$ to $z = 7.33$. The side and projected lengths are 10 kpc for both projections. The black circles represent the location of the identified clumps.
    We set $t=0$ at $z = 7.70$, when the galaxy consists of four luminous clumps.}
    \label{fig:time_evolution}
\end{figure*}

\subsection{Clump formation induced by merger}\label{subsec:clump_formation}
In this section, we study the formation of clumps during a merger. The top panels of Figure \ref{fig:time_evolution} show the time evolution of dark matter distribution around FL957, and the bottom panels represent gas density distribution in the same central region of 10 kpc. 
At $z = 8.01$, we see two dark matter halos, one in the center of the panel and the other one in the right bottom.
The two galaxies experience the first pericenter passage at $z=7.85$, which enhances the tidal compression of the surrounding gas and forms dense clumps. After the first passage ($z=7.70$), we identify four clumps and this number is larger than that of merging halos.
These clumps are formed off-center from the halos. Tail structures coming from clumps 
are formed by tidal effects during the galaxy interaction
\footnote{The movie of clump formation is obtained \href{https://www.dropbox.com/scl/fi/y0551b0vydg6jvpx2n7ly/my_animation.gif?rlkey=oj2b6fryqbzma3e9gvf2ex5uh&dl=0}{here}. The movie shows the time evolution of gas (left) and dark matter (right) densities from $z=8.5$ to $z=7.0$.}.
The formation of clumps triggered by galaxy merger has been investigated by isolated simulations which initially set an Antennae-like galaxy merger \citep[e.g.,][]{Renaud:2013, Renaud:2014, Renaud:2015}. These previous studies have found that the first two pericenter passages increase compressive (curl-free) turbulence and lead to clumpy morphologies and bursty star formation with a timescale of 10-30 Myr. Their results are consistent with our findings from cosmological simulations at high redshift. \notice{Unlike the isolated simulations, which set initial parameters for galaxies and their orbits to reproduce the morphology and kinematics of merging galaxies \citep[e.g.,][]{Renaud:2013, Renaud:2015, Fensch:2017, Maji:2017}, }
our cosmological simulations naturally demonstrate the formation of clumps due to mergers occurring at high redshift. Our simulations can also investigate statistics by connecting to a major merger rate density (see section \ref{subsec:number_abundance}).

We estimate the mass of a perturber that generates clumps based on the 0th-order approach of a spherical collapse model for a homogeneous sphere \citep{Steidel:1998, Overzier:2016}.
A merger-induced clump is formed if the mass is
\begin{equation}
M_{\rm clump} = \bar{\rho}\delta \times V_{\rm clump} \gtrsim M_{\rm min}, \label{eq:clump_formation}
\end{equation}
where $M_{\rm min}$ is the minimum baryon mass of the observable clump we define. Here, $\bar{\rho}, \, \delta, V_{\rm clump}$ are the mean gas density around the progenitor, typical overdensity induced by the merger, and volume of the formed clump. Assuming that the perturber compresses the surrounding gas contained in a volume with its size $\sim R_{\rm per}$, the formed clump mass is re-written as 
\begin{equation}
    M_{\rm clump} = 2\times 10^8 \, M_\odot \, \left(\frac{\bar{\rho} \delta}{250 \, {\rm cm^{-3}}}\right) \left(\frac{R_{\rm  per}}{200\, \rm pc}\right)^{3}. \label{eq:clump_mass}
\end{equation} 
We normalize the relevant quantities by typical values found in our simulations (Table \ref{table:clump_properties}). For example, $\bar{\rho}\delta \sim 250\, {\rm cm^{-3}}$. The minimum mass of a clump that emits \OIII lines bright enough to be observed is derived based on the young stellar mass criteria $M_{*{\rm (young)}, {\rm min}} = 10^7\, M_\odot$ in section \ref{subsec:parameter_determination}, 
\begin{equation}
    M_{\rm min} \sim \frac{M_{*{\rm (young)}, {\rm min}}}{1- f_{\rm gas}} = 10^8 \, M_\odot \left(\frac{1-f_{\rm gas}}{0.1}\right)^{-1},  \label{eq:critical_mass}
\end{equation}
where $f_{\rm gas}$ is the gas fraction defined in section \ref{subsec:clump_properties}. Young clumps have $f_{\rm gas} \sim 0.9$ 
(Figure \ref{fig:clump_property_statistics}). Substituting equations (\ref{eq:clump_mass}) and (\ref{eq:critical_mass}) into eq.(\ref{eq:clump_formation}), we obtain the characteristic size of the baryonic ``wake" induced by merger encounter (perturber) as $R_{\rm per}\gtrsim 150 - 200 \, {\rm pc}$.
The mass ratio $q \equiv M_{\rm center}/ M_{\rm per}$ 
largely determines the subsequent morphological evolution and the mode of triggered star formation.  We assume the central one is more massive, i.e., $M_{\rm center} \geq M_{\rm per}$. Considering that the central object in the system has a typical size of $R_{\rm center}\simeq 200-230 \, {\rm pc}$ as listed in Table \ref{table:clump_properties} and assuming that both objects have similar gas densities, we estimate the mass ratio to be $q = (R_{\rm center}/R_{\rm per})^3 = 1-3.6$. We thus argue that bright clumpy galaxies in optical emission lines at $z> 6$ can be formed by a major merger.

\begin{figure}
    \centering
    \includegraphics[width = \linewidth, clip]{\figdir/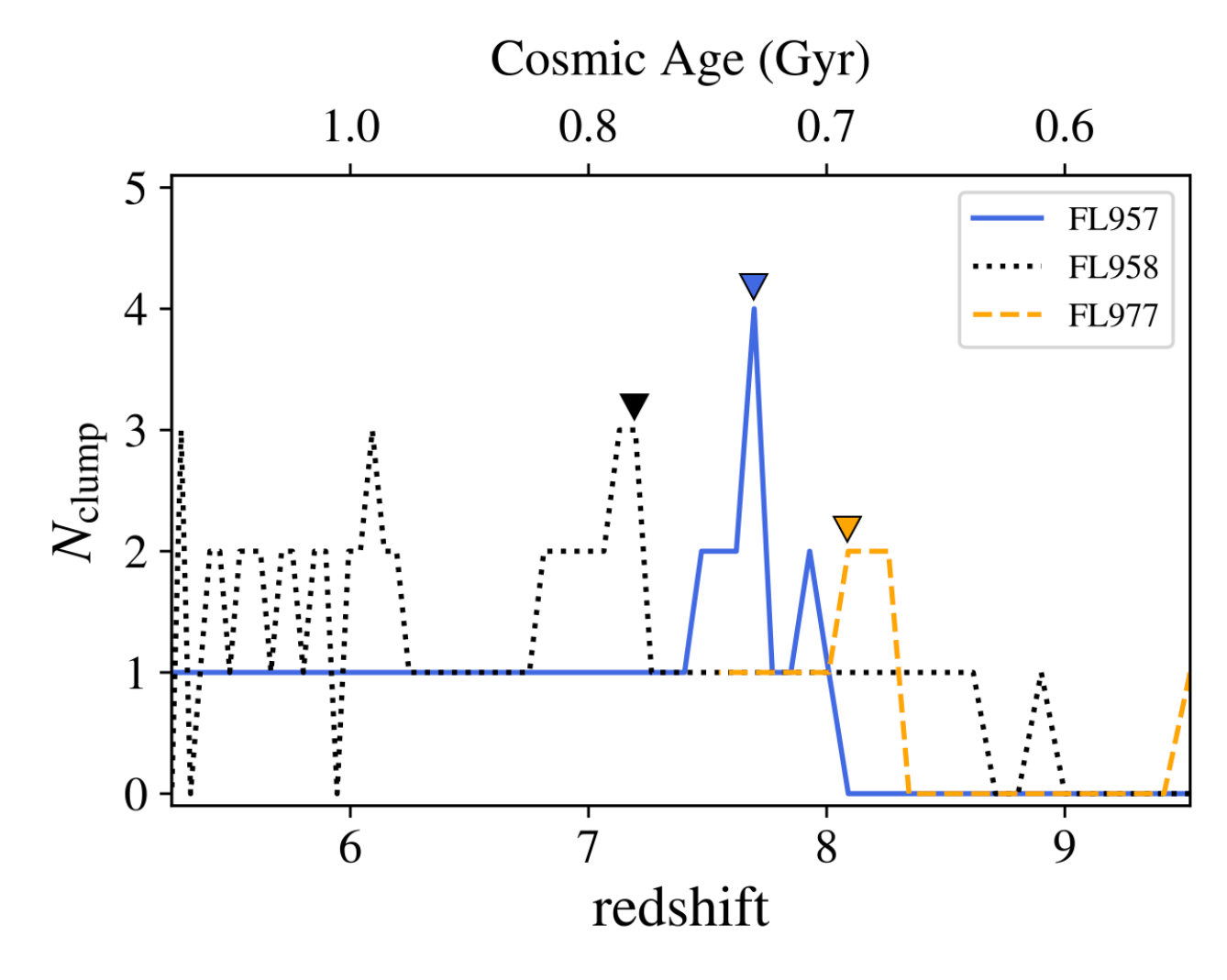}
    \caption{Time evolution of clump number as a function of redshift for the three galaxies shown in Figure \ref{fig:projection}. The inverted triangles represent the redshifts in Figure \ref{fig:projection}. $N_{\rm clump} = 0$ represents that clumps are too faint to be identified by the clump finder.
}
    \label{fig:clump_count}
\end{figure}

\subsection{Clump fate}\label{subsec:clump_fate}
Figure \ref{fig:time_evolution} also follows the subsequent evolution of the clumpy system. A further 18 Myr after the formation of the four clumps, Clump B and C merge with clump A and D resulting in two clumps, which finally merge into a single system 26 Myrs later. 

Figure \ref{fig:clump_count} shows the redshift evolution of the number of clumps for the three galaxies shown in Figure \ref{fig:projection}.  Clumpy structures are observed temporarily and finally merge into the center of the system 30-40 Myr after the multiple systems are formed, whose evolution is seen in other galaxy samples. The typical timescale is the crossing time during the merger:
\begin{equation}
    t_{\rm cross} = \frac{d}{v- v_{\rm center}} 
     \sim 50 \, {\rm Myr} \left(\frac{d}{3 \, {\rm kpc}}\right)\left( \frac{v- v_{\rm center}}{60\, {\rm km/s}}\right )^{-1}, \label{eq:crossing_time}
\end{equation}
where $d$ and $v_{\rm center}$ are the separation and the velocity of the central object. The derived crossing timescale is consistent with the merger timescale in Figure \ref{fig:time_evolution}, and roughly the same order of the orbital (dynamical) timescale of this system in a 10 kpc region, $t_{\rm orb} \sim t_{\rm dyn} \sim 25 \, {\rm Myr}$. Therefore, we see that clumpy systems which are bright in rest-frame optical emission lines are temporary structures and finally merge into one massive galaxy within $\sim$ 50 Myr.

Notice that the two galaxy systems (FL957, FL977) in Figure \ref{fig:clump_count} are identified as $N_{\rm clump} = 0$ at $z\sim 9 $. This implies that these galaxies are too faint to be identified as an \OIII bright object at $z\sim 9$. As noted in Section \ref{sec:method}, our clump identification is based on surface SFR density and the threshold value is 30 $M_\odot \, {\rm yr^{-1}}\, {\rm kpc^{-2}}$ (corresponding to 1 $M_\odot\, {\rm yr^{-1}}$ per clump). However, a lensing effect enables us to identify clumpy features even at $z \gtrsim 9$ as the recent observations report triply lensed galaxies at $z = 10.17$ \citep{Hsiao:2023_MACS0647JD_NIRCam, Hsiao:2023_MACS0647JD_NIRSpec}.

Since our simulations end at $z=5$, we cannot follow the further evolution to the present day.  \notice{Some galaxies form clumps via disk instabilities at $z < 6$, outside the redshift range investigated here.
}
Such clumps still exist at the final snapshot, $z = 5$, but most probably they migrate to the center or they are disrupted due to feedback. Due to our limited resolution, we cannot exclude the possibility that the densest regions of the clumps formed proto-globular clusters that may survive all these events. Future modeling will be required to follow their evolution.


\subsection{number abundance of clumpy galaxies}\label{subsec:number_abundance}
\begin{figure}
    \centering
    \includegraphics[width = \linewidth, clip]{\figdir/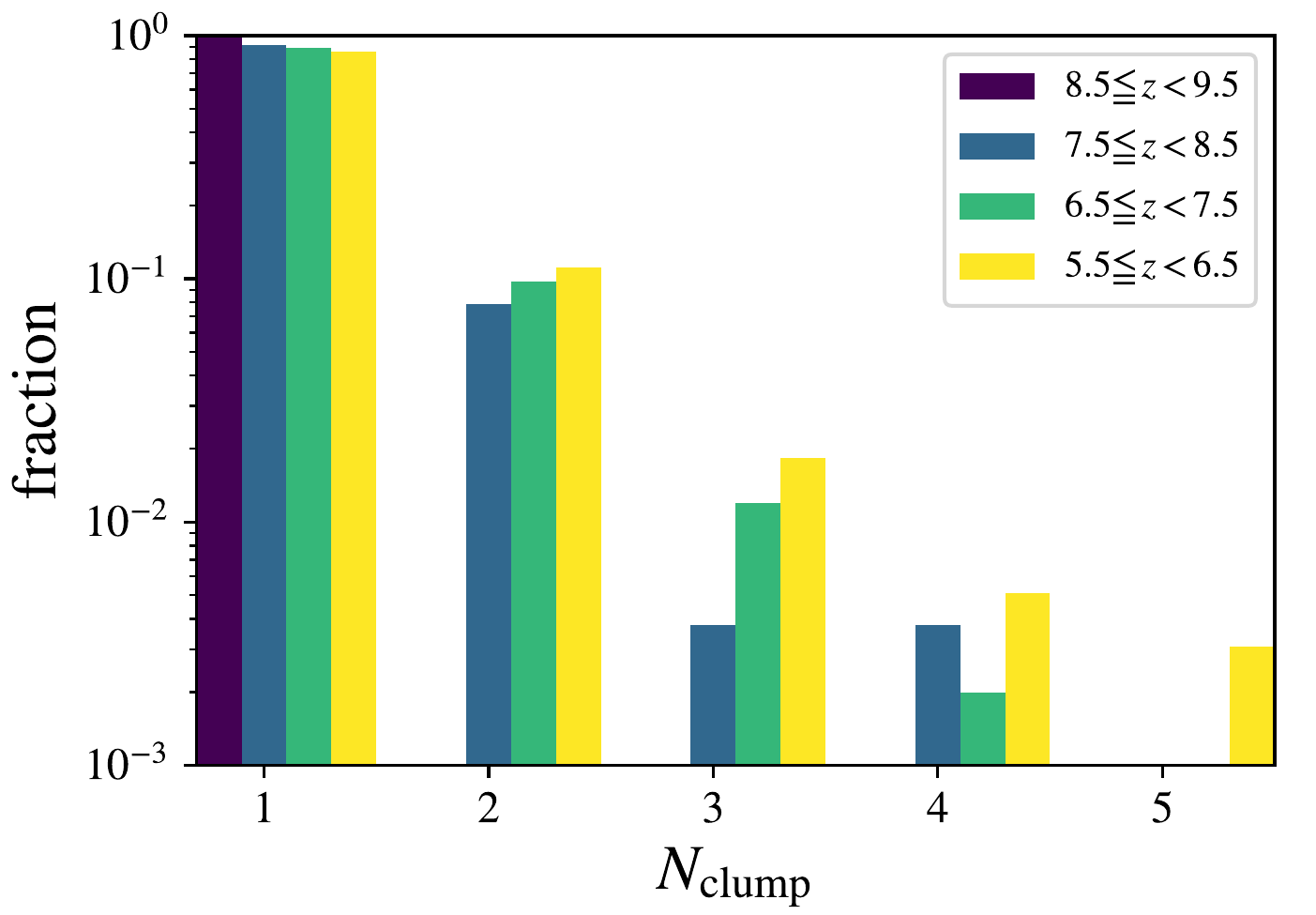}
    \caption{The fraction of each clump number at $z\sim9$ (purple), $z\sim8$ (blue), $z\sim7$ (green), and $z\sim6$ (yellow). The clump number $N_{\rm clump}$ is referred to as the number of clumps in the galaxy system within 10 kpc $\times 10$ kpc region.
}
    \label{fig:clump_fraction}
\end{figure}

Figure \ref{fig:clump_fraction} represents the fraction of clump number at each redshift. Around 90\% of galaxies at $z=8-6$ are observed as a single system ($N_{\rm clump} = 1$). 
From Table \ref{table:FL957_a0p115} and \ref{table:clump_properties}, we see that both observable clumpy galaxies tend to have stellar masses of $M_* \gtrsim 10^9\, M_\odot$ in their systems, even though our clump definition sets a minimum clump mass of $M_* \geq M_{*({\rm young}), {\rm min}} = 10^7 \, M_\odot$. Our sample galaxies at $z \geq 8.5$ have small stellar masses of $M_* \lesssim 10^9 \, M_\odot$, resulting in that all of our samples at $8.5 \leq z \leq 9$ are identified as a single system. 
At these high-redshifts, the corresponding UV magnitude is $M_{\rm UV} \gtrsim -19.7$ \footnote{We use the scaling relation between stellar mass and UV magnitude obtained from \citet{Ceverino:2017, Ceverino:2019} as follows,
\begin{align}
    M_{\rm UV} &= \frac{\log_{10} M_*- 6}{\alpha_*} + M_*^* \,\, (\text{for } z=6-7), \\
    M_{\rm UV} &= \frac{\log_{10} M_*- 6}{\alpha_*} + M_*^* -1 \,\, (\text{for } z=8-9),
\end{align}
where $\alpha_* = -0.394 \pm 0.002$ and $M_*^* = -12.13 \pm 0.03$.
}. Galaxies at $z \lesssim 8$ are bright ($M_{\rm UV} <$ - 19.7) and multiple clumps are identified, which are formed by mergers. This trend is interestingly consistent with the recent JWST observations of \citet{Chen:2023}, who find that luminous $z \simeq 6-8$ galaxies with $M_{\rm UV} < -20.7$ tend to have 3-4 clumps and fainter galaxies with $M_{\rm UV} > -20.7$ tend to be a single system. The fraction of galaxies with three or more clumps ($N_{\rm clump} \geqq 3$) is $\sim$ 0.5 - 1\%. 


We do not find any clumpy systems formed by violent disk instability (VDI) through cold accretion until $z\sim 6$. This is because VDI requires a massive disk with a large disk mass fraction 
\citep{Dekel:2009, Ceverino:2010}. We identify some disk-like galaxies at $z = 8-9$, but the stellar masses are $\lesssim 5 \times 10^9 \, M_\odot$ and have smoothed disk. Specifically, two galaxies with five clumps are seen in a massive system; the total stellar mass in the field is $M_* \sim 10^{10} \, M_\odot$ at $z \sim 5.5$. The clumps in such a massive system are expected to be formed via violent disk instability as same as lower-redshift galaxies \citep{Mandelker:2014, Mandelker:2017, Ceverino:2023, Inoue:2016}. 

\begin{figure}
    \centering
    \includegraphics[width = \linewidth, clip]{\figdir/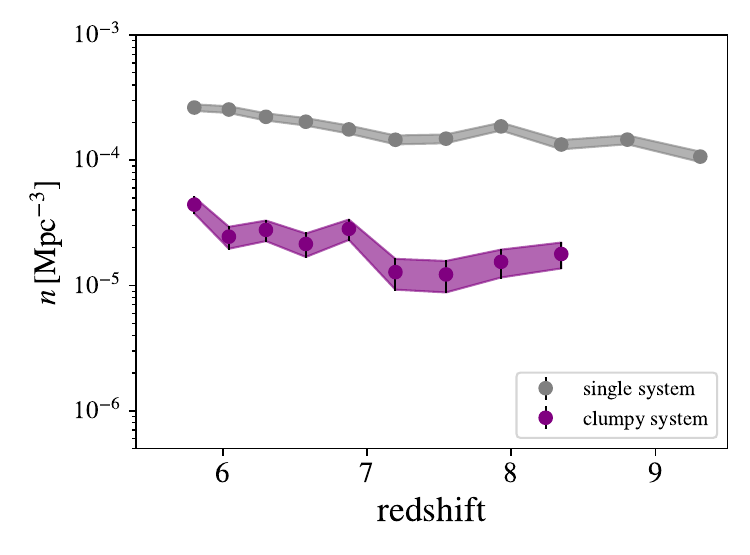}
    \caption{The number density of clumpy galaxies as a function of redshift. The gray and purple plots represent single systems and clumpy systems, respectively.
    The error bars are calculated from Poisson errors, $\sqrt{N}/V_{\rm box}$, where $N$ is the number of clumpy galaxies and $V_{\rm box}$ is the comoving volume of the simulation box.
}
    \label{fig:clump_num_dens}
\end{figure}

\indent
Figure \ref{fig:clump_num_dens} shows the number density of clumpy galaxies. As described in Section \ref{subsec:simulation}, our study employs zoom-in simulations and we need to correct for incompleteness in the number count of clumpy galaxies. We use the results of a larger boxsize $N$-body simulation \citep{Klypin:2011} (refer to Appendix \ref{subsec:abundance_matching} for details). 
The number density of single systems is $(1-3)\times 10^{-4} \, {\rm cMpc^{-3}}$ at $z = 6-9$. We identify all the simulated galaxies at $z=9$ as single systems and the number density is consistent with the UVLF at $M_{\rm UV} \sim - 20$ \citep{Harikane:2023a}. At $z = 6-8$, we estimate the number density to be $(1-5)\times 10^{-5} \, {\rm cMpc^{-3}}$, and this value is enough to be observed by recent JWST surveys. The EIGER survey \citep{Kashino:2023} conducted deep JWST/NIRCam wide-field slitless spectroscopic observations and identified 117 \OIII emitters at $z = 5.33-6.93$ \citep{Matthee:2023}. They found that $\sim$ 1\% of their observed \OIII emitters are closely separated within $< 1"$ (corresponding to $\lesssim$ 6 kpc at $z \sim$ 6). They also derived the \OIII luminosity function at $z\sim 6$. We integrate it within the luminosity range for their samples and obtain the number density of their observed \OIII emitters, $n_{\OIII} = 10^{-2.8} \,{\rm cMpc}^{-3}$. Thus, we can estimate the number density of clumpy systems of their observed \OIII emitters as $\sim 0.01 \times n_{\OIII} = 1.6 \times 10^{-5} \,{\rm cMpc}^{-3}$, consistent with our theoretical abundance of clumpy galaxies at $z = 6-8$. 


We can also derive the merger rate density from the clumpy systems within a 10 kpc $\times$ 10 kpc region following \citet{Mo:2010}. Under the assumption that these pairs will merge on a time scale $\tau_{\rm mrg}$, the merger rate density is related to the clumpy factor $f_{\rm clumpy} \equiv$ (\# of clumpy systems with $N_{\rm clump} \geqq 2$)/(\# of total systems),
according to
\begin{align*}
\dot{n}_{\rm mrg} &= \frac{1}{\langle N_{\rm clump}\rangle} \frac{f_{\rm clumpy} n_{\rm gal}}{\tau_{\rm mrg}} \\
& = 6.3  \times 10^{-5} \, {\rm cMpc^{-3}\, Gyr^{-1}} \\
& \times \left( \frac{f_{\rm clumpy}}{0.1}\right) \left(\frac{n_{\rm gal}}{10^{-4}\, {\rm cMpc^{-3}}}\right) \left(\frac{\tau_{\rm mrg}}{80 \, {\rm Myr}}\right)^{-1},  \stepcounter{equation}\tag{\theequation}
\label{eq:merger_rate}
\end{align*}
where $\langle N_{\rm clump}\rangle$ is the averaged clump number among clumpy systems. The factor 1/$\langle N_{\rm clump} \rangle$ takes into account the fact that a multi-clump system ends in a single merger. Figure \ref{fig:clump_fraction} shows that $\sim$ 80 - 90\% of clumpy systems consist of two clumps, and we thus adopt $\langle N_{\rm clump}\rangle=2$. 
We obtain the clumpy factor from Figure \ref{fig:clump_num_dens} as $f_{\rm clumpy} \sim 0.1-0.2$ and see that the value does not change very much at $z= 6-9$. For the comoving number density of galaxies population $n_{\rm gal}$, we obtain $n_{\rm gal} \sim 10^{-4}\, {\rm cMpc^{-3}}$ with $M_{*} \gtrsim 10^9 M_\odot$ 
\footnote{This stellar mass is for a galaxy system within 10 kpc $\times$ 10kpc, not each clump stellar mass.} from Figure \ref{fig:clump_num_dens}.
The value also does not change largely within a factor of 1-2. We adopt the crossing timescale (eq. (\ref{eq:crossing_time})) as the merger timescale $\tau_{\rm mrg}$. In our case, we take into account a galaxy pair located within a 10 kpc $\times$ 10 kpc region, i.e., the separation of $d \sim 5$ kpc, which gives $\tau_{\rm mrg} \sim 80 \, {\rm Myr}$. The merging timescale is also seen in the top panels of Figure \ref{fig:time_evolution}.
Adopting this merger as a typical case, we obtain the merger rate density of $\dot{n}_{\rm mrg} = 6.3 \times 10^{-5} \, {\rm cMpc^{-3}}\, {\rm Gyr^{-1}}$. This value is roughly consistent with the results from Millennium Simulation at $z > 6$ \citep[fig.2 in][]{Kitzbichler:2008}. We also compare our merger rate density with that of Illustris simulations. \citet{Rodriguez-Gomez:2015} derive the major merger rate $\sim 1 \, {\rm Gyr^{-1}}$ for the descendant stellar masses of $\sim 10^9\, M_\odot$ at $z = 6-9$. From the stellar mass function at $z > 6$ of Illustris simulations \citep{Genel:2014}, the corresponding number density of the descendant mass is $\sim 10^{-4}\, {\rm cMpc^{-3}}$. Therefore, the major merger rate density is $\sim 10^{-4}\, {\rm cMpc^{-3}\, Gyr^{-1}}$, which is in good agreement with our values. These consistencies infer that clumpy galaxies, which are bright in rest-frame optical emission lines, can be linked to major mergers. 

\section{Discussion \& Summary} \label{sec:discussion}
We have studied formation of clumpy galaxies in the epoch of reionization observable in rest-frame optical emission lines.
The clumps we identify are bright in \OIII 5007\AA \,and have radii of $\sim$ 150 - 200\, {\rm pc}, which can be resolved by JWST. We have shown that the observable large clumps are formed by major mergers. The clumps are categorized into two types: proto-bulges dominated by stellar populations older than 50 Myr and off-centered clumps dominated by younger stellar populations. The latter type of clump is formed from gas debris in tidal tails induced by the mergers.
The merger-driven clumps tend to have a high gas fraction of over 90\%, trigger bursty star formation with sSFR $\sim 20-100 \, {\rm Gyr^{-1}}$, and produce young stellar components. The clumpy galaxies account for $\sim$ 10\% of the population, and the fraction remains roughly constant from $z=9$ to $z=6$. We also find that the clumpy systems are short-lived, ending in a merger with a companion(s) within a few tens of Myrs. The number density of clumpy systems is estimated to be $(1-5) \times 10^{-5} \, {\rm cMpc}^{-3}$, which is large enough to be observed in recent JWST surveys \citep{Matthee:2023}.

Formation of clumpy galaxies at $z \lesssim  4$ has been studied in detail \citep{Mandelker:2014, Mandelker:2017, Buck:2017, Ceverino:2023, Inoue:2016}. Especially, \citet{Mandelker:2014} study clump formation in 29 galaxy samples at $z = 1-4$. They define clumps as baryonic overdensities in three dimensions.
They find that 70\% of the clumps are formed \textit{in situ} by violent disk instability while the remaining 30\% are formed \textit{ex situ} by minor mergers. In the present paper, we study clump formation at $z=6-9$ using high-resolution simulations. 
 We define the clumps as \OIII bright objects based on the threshold surface star formation rate so that they can be observed by JWST with a reasonable exposure time. We find that most of the observable clumps are formed by mergers of galaxies with roughly the same baryonic mass. A sufficiently large ``perturber" comparable to the central object with size $\sim$ 200 pc
can generate gas wakes and overdensities to form large, bright clumps. Interestingly, violent disk instability (VDI) is less likely to occur in low-mass galaxies because the disk mass itself is small \citep{Dekel:2009}. 
 In the VDI case, even when it occurs, clumps contain only 1-2 \% of the mass of the central object \citep{Ceverino:2010}, rendering them too faint to be detected through current observations. 
 Our result suggests a change of the mode of clump formation from the predominantly merger-driven scenario in the early universe to VDI at low redshifts at $z \lesssim 4-5$.
 
There are still some caveats in our study. Firstly, we define clumps by adopting a constant threshold of surface star formation rate density. The constant value is motivated to identify observable clumps, but at the same time, we might miss clumps in low-mass galaxies at $z \sim 9$. Such unidentified low-mass clumps might correspond to proto-globular clusters. Recent JWST photometric observations in gravitational lens fields have found young stellar cluster candidates with $M_* = 10^5 - 10^6 \, M_\odot$ at $z = 6- 10$ \citep{Vanzella:2022, Vanzella:2023, Adamo:2024, Bradley:2024, Mowla:2024}. The star clusters are also very small in size with effective radii of less than 1 pc \citep{Adamo:2024}, which our simulations cannot resolve. Some cosmological simulations focus on the formation of dwarf galaxies and stellar clusters at $z > 6$ and have sub-pc scale resolution \citep{Ma:2018_morphologies_and_sizes, Ma:2018_SMF_LF, Calura:2022, Garcia:2023, Sugimura:2024}, allowing one to study the formation of \notice{smaller-scale structures.}

Second, our clump identification is based on rest-frame optical emission lines. Statistics of high-$z$ $(z > 6)$ clumpy galaxies have been conducted from rest-frame UV observation by HST \citep[e.g.,][]{Shibuya:2016, Bowler:2017, Bowler:2022, Mestric:2022}. Some samples have also been observed by ALMA, and they have offsets between FIR emission regions and UV emission regions \citep{Carniani:2017}. Recent JWST multi-band photometry or IFU observations enable us to \notice{perform multiwavelength studies of clumpy galaxies, from the rest-UV to NIR.} The number of clumps is often reported to have spatial offsets between UV, optical, and NIR wavelengths \citep{Colina:2023, Bik:2023, Kalita:2024, Rodighero:2024}, while \citet{Treu:2023} find no dramatic morphology changes across multi-bands. We note that multi-component observations at high redshift ($z > 6$) are \notice{more challenging than at lower redshift \citep[e.g. $z\sim 2-3$, ][]{Mestric:2022, Zanella:2024}, where galaxy morphologies are most of the times less disturbed and present clearer features.}

Our study focuses on young giant clumps that are bright in [O{\sc iii}] because rest-optical emission lines are less affected by dust attenuation than UV emission. These lines trace dense star-forming regions that are gravitationally bound and exist as physical components \citep{Claeyssens:2023, Matthee:2023}. Clumps identified in UV-bands are expected to be heavily attenuated by dust in dense clouds of $n_{\rm gas} \geq 100 \, {\rm cm^{-3}}$ (Figure \ref{fig:projection}). To detect UV clumps in the simulated galaxies and compare them with the current multiwavelength observations, we need to conduct post-processed three-dimensional radiative transfer calculation \citep[e.g., SKIRT, ][]{Baes:2011, Camps_Baes:2020}. We leave it as future studies (Nakazato et al. 2024 in prep. )


 We have found that a clumpy system contains different stellar/clump populations: old ($> 50 \, {\rm Myr}$) clumps at the center of dark matter halos, and young ($< 50 \, {\rm Myr}$) off-centered clumps. The latter type of clumps originates from dense gas clouds that are formed in tidal tails during mergers. Observationally, it would be interesting to identify old clumps with ages of a few hundred Myrs. As shown in Figure \ref{fig:projection}, there are spatial offsets between the young ($\lesssim 50 \,{\rm Myr}$) and old ($\sim 300 \,{\rm Myr}$) components. Those old components are not detected as a clump based on our criteria. Recent JWST observations show that some clumps are detected in rest-frame optical emission lines and continuum, while others are detected only in continuum \citep[e.g.,][]{Hashimoto:2023}. 
 Detailed radiative transfer calculations would also enable us to understand the cause of the spatial offsets between young stellar populations and dust continuum emitting regions as seen in recent observations at $z > 6$ by ALMA \citep{Tamura:2023}.

Emission line flux ratios provide crucial information on clump properties such as the ISM electron density, metallicity, and ionization parameters. For instance, \citet{Yang:2023} use a zoom-in simulation to show that the ratio of $F_{\OIII 5007\, \mathrm{\mathring{A}}}/ F_{\OII 3727\, \mathrm{\mathring{A}}} $ can be used to trace stellar populations younger than several Myr. The ratio is taken by some NIRSpec IFU/NIRISS observations to evaluate the ionization state and metalicity gradients in each pixel \citep{Arribas:2023, Saxena:2024, Tripodi:2024, Rodriguez_Del_Pino:2024, Wang:2022, Venturi:2024}. 
Line ratios are also measured for $z\sim 4$ lensed galaxies to determine and produce resolved dust-to-gas ratio maps \citep{Birkin:2023}. Applying the same method to clumpy galaxies at higher redshift will enable us to characterize young and old clumps and to test the merger-driven scenario we propose here.


\section*{Acknowledgements}
We would like to thank the anonymous referee for very useful comments.
We are grateful to thank Takatoshi Shibuya, Takuya Hashimoto, Yuma Sugahara, Javier \'{A}lvarez-M\'{a}rquez, Luis Colina, Luca Constantin, Carmen Blanco Prieto for the fruitful discussion.
 This work made use of v2.3 of the Binary Population and Spectral Synthesis (BPASS) models as described in \citet{Byrne:2022} and \citet{Stanway:2018}. 
The authors thankfully acknowledge the computer resources at MareNostrum and the technical support provided by the Barcelona Supercomputing Center (RES-AECT-2020-3-0019).
Numerical analyses were carried out on the analysis servers at Center for Computational Astrophysics, National Astronomical Observatory of Japan.
YN acknowledges funding from JSPS KAKENHI Grant Number 23KJ0728 and a JSR fellowship.
DC is a Ramon-Cajal Researcher and is supported by the Ministerio de Ciencia, Innovaci\'{o}n y Universidades (MICIU/FEDER) under research grant PID2021-122603NB-C21. YN and NY acknowledge support from
JSPS International Leading Research 23K20035.

\appendix
\section{Abundance matching} \label{subsec:abundance_matching}
\begin{figure}
    \centering
    \includegraphics[width = 0.55\linewidth, clip]{\figdir/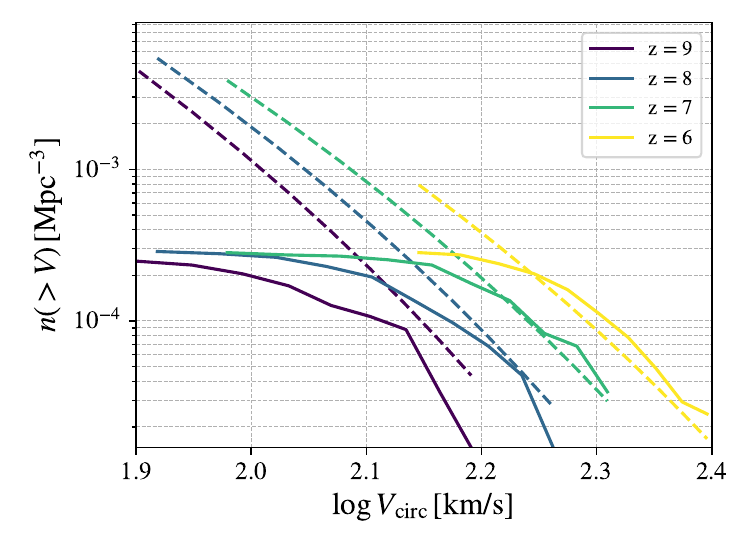}
    \caption{Velocity function of halos at $z=$9(purple), 8(blue), 7(green), and 6(yellow). Solid lines are obtained from FirstLight 62 samples, and dashed lines are from the Bolshoi simulation using the same cosmological parameter \citep{Klypin:2011}. 
}
    \label{fig:velocity_function}
\end{figure}

FirstLight simulation provides a velocity-selected sample of massive galaxies with $\log V_{\rm circ}\,[\mathrm{km/s}] \geq 2.3$ at $z = 5$. The output of the zoom-in simulations contains the galaxies and their progenitors
but not other smaller galaxies in a cosmological volume.
In order to estimate the comoving number density of low mass galaxies, we use the circular velocity function of Bolshoi $N$-body simulation, which has a larger volume of $(250\, h^{-1}\mathrm{cMpc})^3$ \citep{Klypin:2011, Klypin:2016, Rodriguez-Puebla:2016}. 
Figure \ref{fig:velocity_function} compares
the velocity functions at $z=9-6$ obtained from FirstLight (solid lines) with
those from the Bolshoi simulation (dashed lines). 
The latter can be well approximated by 
\begin{equation}
    n (> V) = A V^{-3} \exp \left( - \left[ \frac{V}{V_0}\right]^\alpha \right), 
\end{equation}
where
\begin{align}
    A &= 1.52 \times 10^4 \, \sigma_8^{-3/4}(z) \, (h^{-1}\, {\rm Mpc/km s^{-1}})^{-3}, \\
    \alpha &= 1 + 2.15 \, \sigma_8^{4/3}(z), \\
    V_0 &= 3300 \frac{\sigma_8^2(z)}{1 + 2.5 \sigma_8^2(z)} \, {\rm km s^{-1}},
\end{align}
as derived in \cite{Klypin:2011}. 

We reconstruct the halo and galaxy abundances in the following manner.
We configure 20 velocity bins logarithmically scaled. In each bin, we calculate the ratio of the number density of the Bolshoi and FirstLight simulations, $R_i = n(> V_{\rm circ})_{, {\rm Bolshoi}}/n(> V_{\rm circ})_{, {\rm FL}}$.
For each snapshot subscripted $i$, we compute the circular velocity ($V_{{\rm circ}, i}$) of the galaxy and count the total number of galaxies within a velocity bin. We then use the ratio $R_i$ as a weighting factor to derive the ``true" number of single (clumpy) systems at a given redshift as
\begin{equation}
    N_{\rm single (clumpy)}(z_{\rm bin}) = \sum_{i}^{\# {\rm snap}(z_{\rm bin})} \Theta R_i, 
\end{equation}
where $\Theta$ is set to 1 when snapshot $i$ is single (clumpy) system and otherwise 0. Here $\# {\rm snap}(z_{\rm bin})$ represents the number of snapshots in each redshift bin. Dividing the number by the volume of the FirstLight simulation, i.e., $(40\, {\rm cMpc}/h)^3$, we obtain the number density of the single or clumpy systems at each redshift in Figure \ref{fig:clump_num_dens}.

\section{Numerical Resolution and Pressure Floor} \label{subsec:artificial_pressure_floor}

\notice{
In our simulations, we impose a pressure floor on the condensing gas in galaxies to prevent artificial fragmentation.
It is known that fragmentation can occur if the local Jeans length $\lambda_{\rm J}$ is not well resolved by computational grids with a cell size $\Delta x_{\rm min}$. 
\citet{Truelove:1997} show that artificial fragmentation can be prevented if the Jeans length is always resolved by at least four resolution elements (i.e., $N_{\rm J} \equiv \lambda_{\rm J}/\Delta x_{\rm min}= 4 $). \citet{Ceverino:2010} suggest a more stringent criteria of $N_{\rm J} =7$ by running long-term cosmological simulations.} 

\notice{
In our simulations, 
we set an effective pressure floor at 
\begin{align}
    P_{\rm floor} = \frac{G\rho_{\rm gas}^2 N_{\rm J}^2 \Delta x_{\rm min}^2}{\pi \gamma}, \label{eq:pressure_floor}
\end{align}
with $N_{\rm J} =7$.  We assume a ratio of specific heats of $\gamma = 5/3$ for monoatomic gas and $\rho_{\rm gas}$ is the gas mass density. The pressure in the Euler equation is replaced by eq. (\ref{eq:pressure_floor}) when its pure thermal pressure contribution has a lower value. Artificial fragmentation is effectively prevented by implementing the pressure floor, which is meant to model crudely non-thermal contributions such as local turbulence in the ISM.}

\begin{figure}
    \centering
    \includegraphics[width = 0.55\linewidth, clip]{\figdir/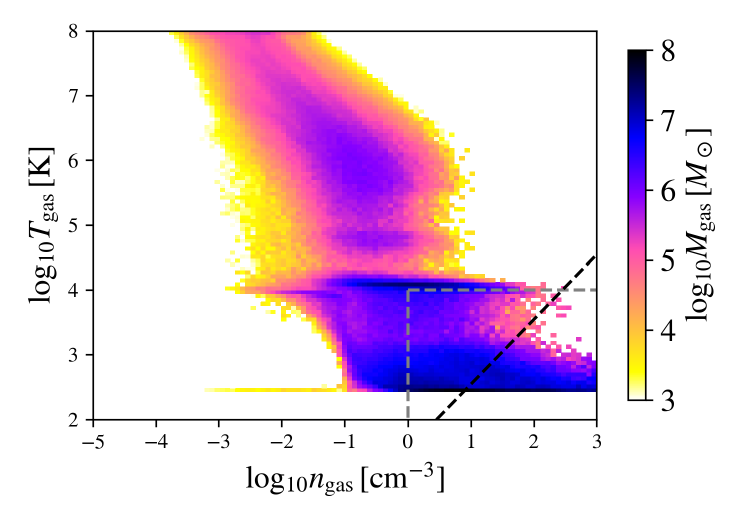}
    \caption{Temperature-density phase diagram for a clumpy system FL958 at $z = 7.2$. The color bar shows the gas mass within each bin $(\Delta \log_{10}n_{\rm gas}, \,\Delta \log_{10}T_{\rm gas}) = (0.08, 0.06)$. Gray dashed lines show the temperature threshold (horizontal) and density threshold (vertical) for star formation. The black dotted line shows the threshold of eq. (\ref{eq:T_J}). We plot all of the gas cells inside a cube with a side length of 10 kpc.
}
    \label{fig:rho_T}
\end{figure}

\notice{
Figure \ref{fig:rho_T} shows a phase diagram for a snapshot of a clumpy system at $z = 7.2$
(see Figure \ref{fig:projection}). Gray dashed lines show the temperature threshold (horizontal, $T_{\rm crit} = 10^4 \, {\rm K}$) and density threshold (vertical, $n_{\rm crit} = 1 \, {\rm cm^{-3}}$) for star formation.
The black dashed line in the left panel shows the following relationship calculated from eq. (\ref{eq:pressure_floor}), 
\begin{align}
    T_{\rm J} = \frac{m_{\rm p}}{k_{\rm B}}\frac{P_{\rm floor}}{\rho_{\rm gas}} 
              = 35 \, {\rm K} \left(\frac{n_{\rm gas}}{1\, {\rm cm^{-3}}}\right) \left(\frac{N_{\rm J}}{7}\right)^2 \left(\frac{\Delta x_{\rm min}}{17\, {\rm pc}}\right)^2, \label{eq:T_J}
\end{align}
where $m_{\rm p}$, $k_{\rm B}$ and $n_{\rm gas}$ are proton mass, Boltzmann constant, and gas number density, respectively. Note that there are gas cells below the black dashed line, on which the pressure floor is imposed in the simulation. The pressure floor has already been implemented in \citet{Ceverino:2010} and detailed analysis on clumps at lower redshifts is given in some previous studies \citep{Mandelker:2014, Mandelker:2017, Inoue:2016, Inoue:2019, Ceverino:2023}
that use simulations with essentially the same set up.}

\bibliography{FL_clumpy}{}
\bibliographystyle{aasjournal}

\end{document}